\documentclass[aps,prb,twocolumn,superscriptaddress,nobibnotes,amsmath,amssymb,reprint]{revtex4-2}

\usepackage{graphicx}
\usepackage{xcolor}
\usepackage[normalem]{ulem}
\usepackage{natbib}
\newcommand{\yrs}[0]{YbRh$_2$Si$_2$}
\newcommand{\myq}[2]{$#1\,\text{#2}$}

\begin{document}
\title{What can we learn from nonequilibrium response of a strange metal?}
\author{B.A.~Polyak}
\affiliation{Osipyan Institute of Solid State Physics RAS, 142432 Chernogolovka, Russian Federation}
\author{V.S.~Khrapai}
\affiliation{Osipyan Institute of Solid State Physics RAS, 142432 Chernogolovka, Russian Federation}
\affiliation{National Research University Higher School of Economics, 20 Myasnitskaya Street, Moscow 101000, Russian Federation}
\author{E.S.~Tikhonov}
\email[e-mail:]{tikhonov@issp.ac.ru}
\affiliation{Osipyan Institute of Solid State Physics RAS, 142432 Chernogolovka, Russian Federation}
\affiliation{Laboratory for Condensed Matter Physics, HSE University, Moscow, 101000 Russia}
\begin{abstract}
We critically address the recent experiment~\cite{Chen2023} on nonequilibrium transport and noise in a strange metal~\yrs{} patterned into the nanowire shape. In the long device, resistivity, differential resistance and current noise data seem to be consistent allowing us to extract electron-phonon coupling and the temperature dependence of electron-phonon scattering length. The obtained values can be reconciled with the experimental data for the short device only assuming the significant contact resistance. We discuss its possible origin as due to the current redistribution between \yrs{} and its gold covering, and reveal that this redistribution contact resistance should be proportional to the \yrs{}~resistivity. We also discuss some subtleties of the noise measurements. Overall, neglecting electron-phonon energy relaxation even in the shortest devices is arguable so that the observed shot noise suppression can hardly be attributed to the failure of quasiparticle concept.
\end{abstract}
\maketitle
The origin of strange metal behavior which is manifested in the linear temperature~($T$) dependence of resistivity down to lowest~$T$ in some materials~\cite{Lohneysen1994,Fournier1998,Legros2019,Nguyen2021,Jaoui2022}, remains without generally accepted theoretical explanation~\cite{Shaginyan2013,Hwang2019,Patel2019,Shaginyan2019,Volovik2019,Sadovskii2020,Sadovskii2021}. On the experimental side, beyond common resistivity measurements novel approaches~\cite{Nakajima2020,Michon2023,Chen2023} are required. In particular, the recent paper by Liyang Chen et al.~\cite{Chen2023} reports on the measurements of shot noise in the heavy fermion strange metal~\yrs{} patterned into the nanowire shape. The authors claim that the observed shot noise suppression can not be attributed to the electron-phonon energy relaxation in a standard Fermi liquid model but rather indicates the failure of quasiparticle concept. This interpretation has been criticized~\cite{Shaginyan2023,Polyak2024} which motivated us to discuss in the present manuscript the peculiarities of the nonequilibrium transport approach to the study of strange metals. 

\begin{figure}[h]
\begin{center}
\includegraphics[width=\columnwidth]{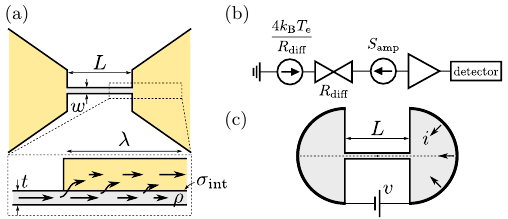}
\end{center}
\caption{Device geometry and measurement setup. (a)~Schematic representation of a device. \yrs{} film $t=60\,\text{nm}$ thick (gray) is patterned into a constriction connecting two pads. These pads are additionally covered with \myq{200}{nm} of gold (yellow). The length and the width of the constriction are~$L$ and~$w$, respectively. Interface between \yrs{} and gold is characterized by conductivity~$\sigma_{\text{int}}$, current transfer length is~$\lambda$. (b)~Geometry used in numerical modelling. We calculate the total current through the constriction, $i$, in response to the applied bias voltage, $v$. (c)~Equivalent circuit of noise measurement setup.}
\label{fig1}
\end{figure}

Fig.~\ref{fig1}(a) demonstrates the sketch of studied devices. Patterned from a~$t=60\,\text{nm}$ thick \yrs{} film (gray) with resistivity~$\rho$, grown on germanium substrate, they are represented by nanowire-shaped constrictions of length~$L$ and width~$w$, connected to source and drain pads. As shown by yellow shading, these two pads are additionally covered with~\myq{200}{nm} of gold. We note that at low bath temperature ($T_0$) the gold conductivity is approximately $10$~times higher than that of \yrs{}~(at~\myq{3}{K}). Bottom part of Fig.~\ref{fig1}(a) shows schematically the current density vector both in~\yrs{} and in gold covering in the region where current redistributes between the two layers. This region extends for the so-called current transfer length~$\lambda$ depending on the quality of the \yrs/Au interface which we quantify with the interface conductance per unit area,~$\sigma_{\text{int}}$. Overall, the authors provide data for three devices with short constrictions, $L\lesssim 1\,\mu\text{m}$, further refered to as short devices, and for one device with long constriction, $L=28\,\mu\text{m}$, further refered to as long device. The widths of all nanowires range from \myq{140}{nm} to~\myq{300}{nm}. We estimate geometric dimensions of all constrictions from the available scanning electron micrograph images. As we argue in this manuscript, geometry of the devices may require taking into account the contribution to the measured resistance and noise not only from the nanowires themselves but also from the significant part of the pads. 

\begin{figure}[h]
\begin{center}
\includegraphics[width=\columnwidth]{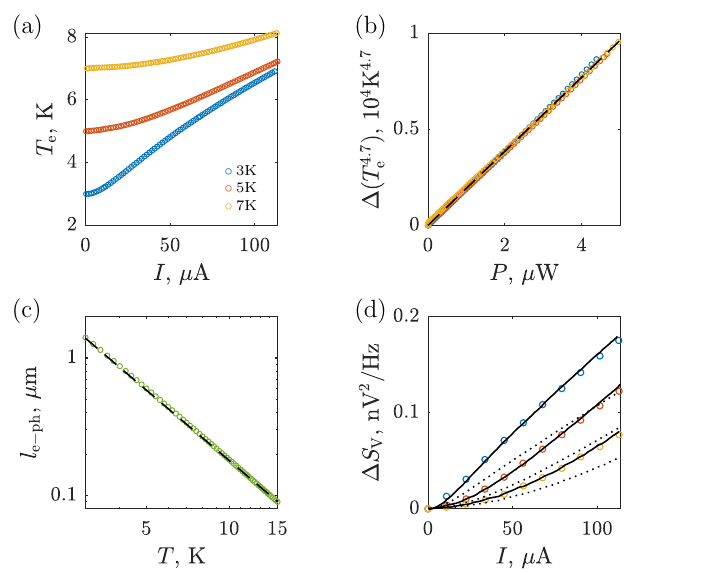}
\end{center}
\caption{Nonequilibrium response of the long nanowire.
(a)~Electronic temperature of the nanowire as a function of bias current. (b)~Electron-phonon power flow obeys $P=\Sigma_{\text{e-ph}}(T_{\text{e}}^{4.7}-T_{\text{ph}}^{4.7})$ with $\Sigma_{\text{e-ph}}=9.6\cdot10^8\,\text{W}/\text{K}^{4.7}\text{m}^3$. (c)~Electron-phonon scattering length as a function of temperature. Dashed line is~$l_{\text{e-ph}}\propto T^{-1.7}$. (d)~Voltage noise spectral density of the device. Symbols represent the data from the experiment~\cite{Chen2023}. Both dotted and dashed lines correspond to $\Sigma_{\text{e-ph}}=9.6\cdot10^8\,\text{W}/\text{K}^{4.7}\text{m}^3$ but with $S_{\text{amp}}=0$ and $S_{\text{amp}}=5\cdot10^{-25}\,\text{A}^2/\text{Hz}$, respectively.}
\label{fig2}
\end{figure}

We start our discussion with the analysis of experimental data for the long device. Here, the central constriction is approximately $30$ times longer than that for three short devices so that the possible effect of the interface may be most likely neglected. Below we show that electron-phonon scattering length in \yrs{} is~$l_{\text{e-ph}}(3\,\text{K})\approx1\,\mu\text{m}$ and decreases with increasing temperature. This ensures that besides short regions near the pads, in the presence of bias current~$I$ electron system in the long constriction is described by position-independent electronic temperature~$T_{\text{e}}(I)$. At all~$T_0$ this dependence can be obtained using a standard procedure. From the differential resistance data provided in Supplementary Materials~(SM) Fig.S3B~\cite{Chen2023}, we extract $R=V/I$ at $T_0=3\,\text{K}$, $5\,\text{K}$ and \myq{7}{K}. Further, attributing the growth of $R(I)$ with increasing current to the increase of~$T_{\text{e}}$, we extract $T_{\text{e}}(I)$. Here, we use the fact that the $T$-dependence of the normalized resistance of the devices is demonstrated to be the same as that for the unpatterned film, see Fig.1C~\cite{Chen2023}. The obtained curves $T_{\text{e}}(I)$ are shown in~Fig.~\ref{fig2}(a). Without noise measurements, these data already allow one to estimate electron-phonon coupling which describes the power flow from electron to phonon subsystem via 
\begin{equation*}
P_{\text{e-ph}}=V\Sigma_{\text{e-ph}}(T_{\text{e}}^n-T_{\text{ph}}^n),
\end{equation*}
where $V$ is the system volume and the exponent~$n$ typically varies in the range~$n\approx3-5$~\cite{Giazotto2006}. The devices are patterned on crystalline germanium substrates ensuring $T_{\text{ph}}=T_0$. In a steady state $P_{\text{e-ph}}=P_{\text{J}}$, where $P_{\text{J}}$ is the released Joule heat power so that  
\begin{equation}
P_{\text{J}}=V\Sigma_{\text{e-ph}}(T_{\text{e}}^n-T_{0}^n).
\label{ephrelax}
\end{equation}
In Fig.~\ref{fig2}(b) we show that $n=4.7$ fits the data perfectly with $\Sigma_{\text{e-ph}}=9.6\cdot10^8\,\text{W}/\text{K}^{4.7}\text{m}^3$ (we take $w=300\,\text{nm}$). From here, we extract the $T$-dependence of the electron-phonon scattering length using~\cite{Denisov2020a}
\begin{equation*}
l_{\text{e-ph}}=L\left[{\cal L}/nT^{n-2}\Sigma_{\text{e-ph}}R(T)\right]^{1/2},
\end{equation*}
where ${\cal L}=2.44\cdot10^{-8}\,\text{W}\Omega\text{K}^{-2}$ is the Lorenz number. The result is shown in Fig.~\ref{fig2}(c) and in the given temperature range can be reasonably approximated as $l_{\text{e-ph}}\propto T^{-1.7}$, see the dashed line. Importantly, $l_{\text{e-ph}}(3\,\text{K})\approx1.4\,\mu\text{m}$ ensuring the possibility to introduce position-independent electronic temperature~$T_{\text{e}}(I)$ which will be further used in noise treatment. 

Before discussing nonequilibrium noise of the long device, we note some general details of the noise measurements. In the setup used in~\cite{Chen2023}, see Fig.~\ref{fig1}(b) for the schematic circuit, the voltage noise before amplification is determined by two contributions,
\begin{equation}
S_{\text{V}}(I)=\left[\frac{4k_{\text{B}}T_{\text{e}}(I)}{R_{\text{diff}}(I)}+S_{\text{amp}}\right]R_{\text{diff}}^2(I),
\label{noiseexpectation}
\end{equation}
where the first term comes from the current noise of the device itself, the second term is defined by the parasitic input current noise of the amplifiers and $S_{\text{V}}(0)$ is the voltage noise in equilibrium. Note that in the nonlinear regime $R_{\text{diff}}$ depends on~$I$ so that $S_{\text{amp}}$ contributes to the measured excess voltage noise. Whether one can neglect $S_{\text{amp}}$ or not, depends on the interplay between the two terms in~(\ref{noiseexpectation}). Therefore, along with determining the gain it is also the goal of calibration to get the magnitude of amplifier noise. Typically, when one uses homemade voltage amplifier at liquid helium temperature, its input current noise is on the order of $10^{-27}\,\text{A}^2/\text{Hz}$~\cite{Reznikov1999,Tikhonov2014,Muro2016} where the precise value depends on the used transistor, operating frequency and device resistance~\cite{Arakawa2013}. The authors of~\cite{Chen2023} use room-temperature commercial preamplifiers, low-noise LI-75, and SR-560 which is usually not used in noise measurements. One can therefore expect $S_{\text{amp}}\gtrsim10^{-25}\,\text{A}^2/\text{Hz}$. 
We note that unless few orders of magnitude larger, this current noise can not be extracted from the setup calibration discussed in SM~section~$2$~\cite{Chen2023}. Namely, the calibration was performed by detecting the room temperature thermal noise of a variety of resistors, see SM~Fig.S1.
At room temperature, the current noise of a typical used $100\,\Omega$ resistor is $1.6\cdot10^{-22}\,\text{A}^2/\text{Hz}$ and by far exceeds the expected value of $S_{\text{amp}}$ so that the presented calibration procedure is absolutely helpless in its determination. At the same time, the current noise of a strange metal long device cooled down to~\myq{5}{K} and with a resistance of approximately~$300\,\Omega$ is $9\cdot10^{-25}\,\text{A}^2/\text{Hz}$ which may easily be comparable to the expected value of $S_{\text{amp}}$. In other words, the setup calibration is performed for the values of current noise which are two orders of magnitude greater than those utilized in the experiment.

The value of $S_{\text{amp}}$ in the experiment can be estimated by comparison of experimental results for the voltage fluctuations of the long device, see SM~Fig.S3A of~\cite{Chen2023}, to what one can expect based on the obtained curves~$T_{\text{e}}(I)$ presented in~Fig.~\ref{fig2}(a). By solid lines in~Fig.~\ref{fig2}(d) we show the best fits to the experimental data (symbols) of~\cite{Chen2023}. These fits are obtained using~(\ref{noiseexpectation}) with $S_{\text{amp}}=5\cdot10^{-25}\,\text{A}^2/\text{Hz}$ which perfectly falls in the above order of magnitude expectation. We emphasize that considering current noise of the device in the form of thermal noise with electronic temperature elevated above bath temperature in~(\ref{noiseexpectation}) is valid in the presence of strong electron-phonon scattering. Dotted lines additionally illustrate the fits obtained with zero preamplifier noise. Note these fits go below the experimental data indicating contribution of the parasitic noise.

As a final remark, we note that the obtained above $T$-independent value of~$\Sigma_{\text{e-ph}}$ implies the temperature dependence of $\Gamma=(e/k_{\text{B}})^2\Sigma_{\text{e-ph}}/\sigma$ since conductivity changes by approximately $40\%$ in the temperature range from~\myq{3}{K} to~\myq{7}{K}. At the same time, the authors of~\cite{Chen2023} used in their fits $T$-independent $\Gamma\approx9\cdot10^9\,\text{K}^{-3}\text{m}^{-2}$, see section~$5$ in~SM. This difference may explain the better quality of our fits for the differential resistance.

We now discuss the short device ($L=660\,\text{nm}$) presented in the main text, see Fig.2A~\cite{Chen2023}. Given the width of the constriction is close to that of the long nanowire, one might have expected its linear response resistance at~\myq{3}{K} to be around~$6\,\Omega$. The actual value is~$36\,\Omega$ demonstrating there is significant inconsistency between formally extracted resistivities of the two devices. Note that additional data on two more short devices presented in~SM~Fig.S5 reveal the similar discrepancy. In principle, this inconsistency can partly be attributed to the nanoscale patterning of the devices using reactive ion etching which inevitably damages the edges of the constrictions~\cite{pc}. Fig.1(C,D) of~\cite{Chen2023} compares temperature and magnetic field~($B$) behavior of the unpatterned film with that of the nanowire-patterned device and demonstrates similar results in terms of normalized resistances. Therefore, while there is approximately $6$-fold difference in the resistivities we believe that the underlying physics is still there so that damaged edges are hardly responsible for the discussed inconsistency. The authors of~\cite{Chen2023} use Fig.1(C,D) to claim that the total resistance of the device is dominated by the constrictions and the pads contribution is negligible. Below we explain that this argument is in fact incorrect and the pads contribution, which we further refer to as contact resistance, can not be excluded based on the observation of similar for the film and for the patterned devices $R(T)$ and $R(B)$~dependences.

To illustrate the idea, we consider the geometry depicted in the bottom part of~Fig.~\ref{fig1}(a). Further, we take gold conductivity to be infinitely large. Upon leaving the constriction, current starts to redistribute between the \yrs{} film and the gold cover. This redistribution stops after going distance~$\lambda$ deep inside the pad along the interface. This distance is called the current transfer length and depends on the film resistivity and on the interface conductivity,~$\sigma_{\text{int}}$. For $\sigma_{\text{int}}=0$ current doesn't flow over the interface and remains completely in the \yrs{} film. In this case, the contact resistance is log-divergent with the contact size~$a$ as $R_{1}(a)=\rho\ln(2a/w)/(\pi t)$. On the other hand, if all voltage drop occurs across the interface between the two layers, which happens for~$\rho=0$, the contact resistance is $R_2(a)=2/(\pi a^2\sigma_{\text{int}})$. In an infinite contact with both~$\rho$ and $\sigma_{\text{int}}$ finite, the reasonable quantitative estimate for both the current transfer length and the contact resistance is obtained from $R_1(\lambda)=R_2(\lambda)\sim R_{\text{cont}}$, that is
\begin{equation*}
\frac{\rho}{\pi t}\ln\left(\frac{2\lambda}{w}\right)=\frac{2}{\pi \lambda^2\sigma_{\text{int}}}\sim R_{\text{cont}}.
\end{equation*}
In Supplemental Material we provide the analytical solution and demonstrate that in the limit of large enough $\lambda\gg w$ the above estimate reproduces the exact result up to only a $10\%$ correction for the logarithm argument. Importantly, the $T$-dependence of~$\lambda$ is weak and closely follows $\lambda\propto \rho^{-1/2}$. Therefore, up to an unimportant log-factor the contact resistance is proportional to the \yrs{} resistivity,
\begin{equation}
R_{\text{cont}}\propto \rho.
\label{ideal}
\end{equation} 
As a result, observation of the identical $T$ and $B$ response for the parent film and for the patterned device doesn't ensure negligible contact resistance. We also note that for the realistic finite-size devices contact resistance may further be enlarged due to the factors not considered in this idealized picture. In Supplemental Material we show that the $T$-dependence of linear-response resistance for the short device presented in the main text, see Fig.2A~\cite{Chen2023}, is reasonably approximated using $\lambda(3\,\text{K})\approx280\,\mu\text{m}$ which is comparable to the pads size and indicates the possible importance of current redistribution effect.

Given the great difference between the expected and the observed resistance of short devices, we argue that the current transfer length may exceed the dimensions of constrictions by orders of magnitude and be comparable to or even exceed the dimensions of devices with pads included. To numerically simulate the differential resistance, we choose the geometry of Fig.~\ref{fig1}(c). Here, the radius of pads equals~$20\,\mu\text{m}$ and the current redistribution on the lateral scale of these pads is neglected. The electrodes are indicated by thick black lines. On the one hand, this geometry is close to the geometry of the patterned film nearby the constriction in real devices. On the other hand, the scale of $20\,\mu\text{m}$ is large enough to capture the nonlinearity of differential resistance, since due to $l_{\text{e-ph}}\lesssim1\,\mu\text{m}$ electronic temperature in the presence of current reaches the value of~$T_0$ on the spatial scale of few micrometers beyond the constriction. Additionally, crucial to the numerical calculations for this highly nonuniform geometry, thus chosen size is not too large to require inadequate computing power.

\begin{figure}[h]
\begin{center}
\includegraphics[width=\columnwidth]{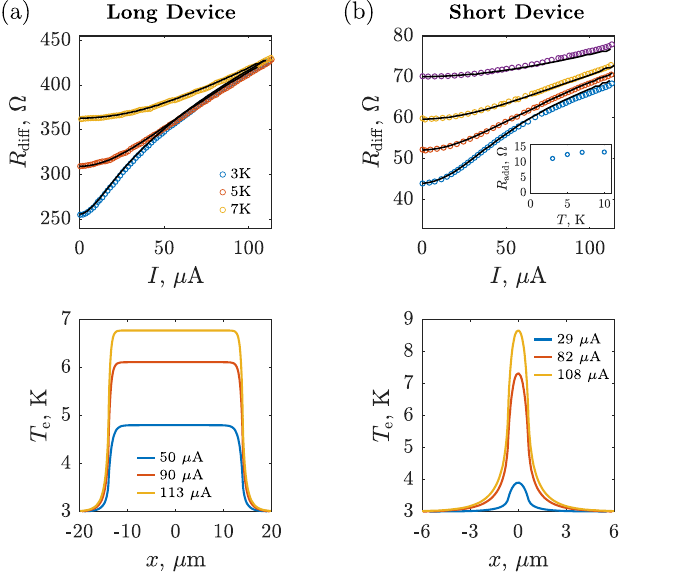}
\end{center}
\caption{Numerical simulation of nonequilibrium response for both (a)~long and (b)~short (device~$\#3$) constrictions. Symbols in top panels reproduce the experimental results of~\cite{Chen2023}. Solid lines are fits obtained with numerical simulation (see text). The inset of top panel in~(b) demonstrates the $T$-dependence of the bias-independent contact resistance. The bottom panels demonstrate the spatial profile of electronic temperature along the dashed line depicted in Fig.~\ref{fig1}(c).}
\label{fig3}
\end{figure}

We first implement our numerical approach for the long device. Symbols in the top panel of Fig.~\ref{fig3}(a) reproduce the experimental data from the SM~Fig.S3B~\cite{Chen2023}. Solid lines are fits obtained with $\Sigma_{\text{e-ph}}=6.1\cdot10^8\,\text{W}/\text{K}^5\text{m}^3$ and $n=5$ in (\ref{ephrelax}) and the $T$-dependence of \yrs{} resistivity $\rho(T)=10.8+1.67\,T\,[\mu\Omega\cdot\text{cm}]$ which captures the linear-response resistance data and fits the data of Fig.1C~\cite{Chen2023} with better than $10\,\%$ accuracy in the temperature range below~$10\,\text{K}$. Note that the difference between thus obtained $\Sigma_{\text{e-ph}}$ and the value extracted from Fig.~\ref{fig2}(b) is due to the slightly different power-law of electron-phonon cooling rate. The spatial profile of electronic temperature along the dashed line depicted in Fig.~\ref{fig1}(c) is demonstrated in the bottom panel of Fig.~\ref{fig3}(a) with $x=0$ corresponding to the center of the constriction.

Having extracted $\Sigma_{\text{e-ph}}$, we attempt to fit the bias dependence of differential resistance for one of short devices. For this purpose we choose the device $\#3$ from SM~Fig.S5E~\cite{Chen2023}. Among the overall presented three short devices, this is the narrowest one so that we expect the contact resistance to be significant enough but not dominating over the constriction resistance compared with two other short devices. Note also that while the constriction of this device is approximately two times narrower than the constriction of the device~$\#2$ from SM~Fig.S5(A-C), its resistance is only~$\approx30\%$ greater which again may indicate the significant contact resistance in all short devices. We take the length and the width of the constriction in device~$\#3$ to be $l=1.3\,\mu\text{m}$ and $w=155\,\text{nm}$, respectively; the corresponding resistivity is taken to be the same as for the long device. Symbols in the top panel of~Fig.~\ref{fig3}(b) reproduce the experimental data from the SM~Fig.S5E~\cite{Chen2023}. Solid lines are fits obtained with
\begin{equation*}
R_{\text{diff}}=\frac{dv}{di}+R_{\text{add}}(T),
\end{equation*}
where $v$ is the voltage applied between two electrodes in Fig.~\ref{fig1}(c), $i$ is the calculated current and $R_{\text{add}}(T)$ is chosen as current-independent quantity to fit the linear-response resistance for all four temperatures of interest. $R_{\text{add}}$~must include, at least, the contribution to contact resistance due to current redistribution across the imperfect \yrs{}/gold interface beyond the semicircular pads depicted in Fig.~\ref{fig1}(c). Additional contribution to~$R_{\text{add}}$ may come from finite gold resistivity and possibly present interface defects. Taking into account only current redistribution effect, in accordance with~(\ref{ideal}) one should expect $R_{\text{add}}\propto \rho$. In the inset we demonstrate the obtained $R_{\text{add}}(T)$~dependence which turns out to be slower than~$\rho(T)$ indicating that current redistribution is not the only effect contributing to the contact resistance. Additionally, in the bottom panel of Fig.~\ref{fig3}(b) we plot the spatial profile of electronic temperature along the short constriction. Both bottom panels of Fig.~\ref{fig3} verify that electron-phonon scattering is strong enough so that the devices can not be considered as in phase-coherent regime which was recently analyzed theoretically~\cite{Nikolaenko2023}.

\begin{figure}[h]
\begin{center}
\includegraphics[width=\columnwidth]{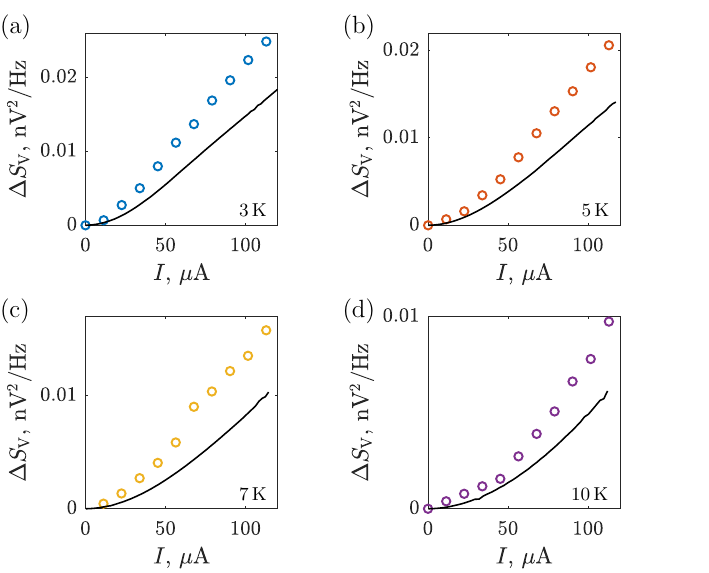}
\end{center}
\caption{Numerical simulation of nonequilibrium noise for the device~$\#3$. Symbols represent the data from the experiment~\cite{Chen2023}. Solid lines are fits calculated using the position-dependent electronic temperature from the bottom panel of~Fig.~\ref{fig3}(b) (see text).}
\label{fig4}
\end{figure}

Finally, we compare the experimental results for nonequilibrium noise in the short device with what we expect based on the spatial temperature profile demonstrated in Fig.~\ref{fig3}(d). By symbols in Fig.~\ref{fig4} we reproduce the data for the device~$\#3$ from SM~Fig.S5F~\cite{Chen2023}. Solid lines are fits calculated using $\Delta S_{\text{V}}=4k_{\text{B}}T_{\text{N}}(dv/di)$. Here, $T_{\text{N}}$ is the noise temperature of the part of the device depicted in Fig.~\ref{fig1}(c) and calculated using~\cite{Piatrusha2017}
\begin{equation*}
T_{\text{N}}=\frac{\int T(x,y)(\mathbf{j}\cdot\mathbf{E})\,dx\,dy}{\int (\mathbf{j}\cdot\mathbf{E})\,dx\,dy},
\end{equation*}
where $\mathbf{j}$ is the current density and $\mathbf{E}$ is the electric field in the given point of a device.
Note that in the above expression for~$\Delta S_{\text{V}}$ we neglect the nonequilibrium noise of device regions beyond those from Fig.~\ref{fig1}(c) as well as the contribution from the amplifier noise due to the device resistance nonlinearity. Importantly, the fits go below the experimental data so that considering the amplifier current noise and the contacts noise may become crucial in interpreting the data.

In conclusion, we discuss the recent strange metal experiment~\cite{Chen2023} and demonstrate that electron-phonon scattering most likely can not be neglected in the presented short devices which makes the statement on the failure of quasiparticle concept arguable. At the same time, experimental data from the long device contain important information about electron-phonon coupling and the $T$-dependence of electron-phonon scattering length which is essential for further transport experiments.

We thank Liyang Chen and Douglas Natelson for valuable comments on the device fabrication details. The work is financially supported by RSF Project No. 22-12-00342.

%

\clearpage
\begin{center}
\textbf{\large Supplemental Material}
\end{center}
\setcounter{equation}{0}
\setcounter{figure}{0}
\setcounter{table}{0}
\setcounter{page}{1}
\makeatletter
\renewcommand{\theequation}{S\arabic{equation}}
\renewcommand{\figurename}{Supplemental Material Fig.}
\renewcommand{\bibnumfmt}[1]{[S#1]}
\renewcommand{\citenumfont}[1]{S#1}

Below we solve a problem of current spreading in a layered two-dimensional geometry with poor interface. The results evidence that the contact resistance scales similar to the resistivity of the parent film. Numerical simulations are performed to understand possible impact of the interface resistance in Ref. [Liyang Chen et al., Science \textbf{382}, 907 (2023)]. The experimental deviations could be expained by such a contact effect.

\section{Layout}

\begin{figure}[ht]
	\includegraphics[width = 0.9\columnwidth]{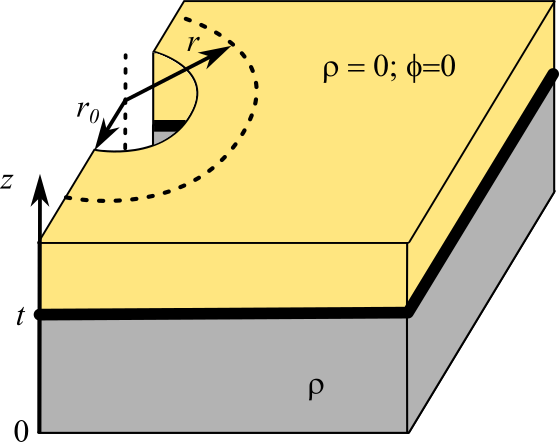}
	\caption{Sketch of a layered contact. Lower grey layer has a finite resistivity (thickness $t$). Upper yellow layer is an ideal metal, which is grounded. A very thin interface between the two layers is depicted by a black thick line. The interface emulates bad contact and in the presence of current there is a finite voltage drop across the interface.}
	\label{fig1SM}
\end{figure}
Fig.~\ref{fig1SM} represents a model of a layered contact. The upper layer is an ideal metal with infinite conductivity, which is grounded, i.e. its electric potential is $\phi = 0$. The lower layer is a metal with finite conductivity $\sigma = 1/\rho$ and thickness $t$. 
The interface between the two metal layers is assumed to have zero thickness and is characterized by a finite conductivity per unit area $\sigma_{i}$. The current inflows the lower layer at the inner radial contact of the radius $r_0$. The center of that inner contact is chosen to be the origin of a cylindrical polar coordinate system $(r,\varphi,z)$. 

\section{Qualitative picture}

A good estimate of the spreading resistance can be achieved with the following argument. Consider the contact of geometry of Fig.~\ref{fig1SM} with the inner and outer radii of $r_0$ and $r_{max}\gg r_0$, respectively. For $\sigma_i=0$ the current does not flow over the interface and remains in the lower layer. In  this case, the contact resistance has a log-divergence in size: $R_1(r_{\text{max}}) = \rho(\pi t)^{-1}\ln{(r_{\text{max}}/r_0)}$. On the other hand, if all the voltage drop occurs across the interface between the two layers, which happens for $\rho=0$, the contact resistance is $R_2(r_{\text{max}})=2(\pi r_{\text{max}}^2\sigma_i)^{-1}$. In an infinite contact with both $\rho$ and $\sigma_i$ non-zero, the current overflows the interface within the region of the radius given by the current transfer length $\lambda$. This length can be estimated from $R_1(\lambda)=R_2(\lambda)$. For such $\lambda$, the contact resistance can be estimated as:
\begin{equation}
	R_c\approx R_1(\lambda) = \frac{\rho}{\pi t}\ln{(\lambda/r_0)}\label{qualitative}
\end{equation}

\section{Solution: half-plane infinite pad}

The spatial distribution of the electric potential is found from a solution of Laplace's equation $\Delta\phi=0$. Owing to the axial symmetry, the potential is independent of $\varphi$ and we find the solution of the form $\phi(r,z) = F(r)G(z)$. It then follows from the Laplace's equation:
\begin{equation}
	\frac{1}{F}\frac{d^2F}{dr^2}+ \frac{1}{rF}\frac{dF}{dr}= - \frac{1}{G}\frac{d^2G}{dz^2}, \label{Laplace}
\end{equation}
which implies that the rhs and lhs of this equation equal the same constant, independent of both $r$ and $z$. This equation is supplemented by the expressions for the current density in the lower metal layer:
\begin{align}
j_z & = -\sigma F(r)\frac{dG}{dz} \label{jz}\\
j_r & = -\sigma G(z)\frac{dF}{dr} \label{jr}
\end{align}
The solution for $G(z)$ is straightforward. Assuming $d^2G/dz^2 = -\gamma^2 G(z)$ and from the boundary condition of $j_z=0$ at the $z=0$ plane, we find $G(z) = G_0\cos{(\gamma z)}$. The boundary condition at the interface $z=t$ is different. We assume that the interface between the two metals has zero thickness and is characterized by a finite conductivity per unit area $\sigma_{i}$. Hence, the finite voltage drop at the interface is connected to the current density along the $z$-axis:
\begin{equation*}
\phi(r,z=t-0)-\phi(r,z=t+0) = \frac{j_z(r,z=t)}{\sigma_{i}},
\end{equation*}
where, by assumption, $\phi(r,z=t+0)=0$. Together with the solution for $G(z)$ and Eq.~(\ref{jz}) this expression determines $\gamma$:
\begin{equation}
	\gamma\tan{(\gamma t) = \frac{\sigma_{i}}{\sigma}}, \label{gamma}
	\end{equation}
which in the limit of $\gamma t\ll 1$ reduces to $\gamma = \sqrt{\sigma_i/(t\sigma)}$. Note that $\gamma^{-1}$ has a meaning of current transfer length, i.e. a typical length on which most of the current from the lower layer flows over to the upper layer. Going back to the Eq.~(\ref{Laplace}), the solution for $F(r)$ is found from:
\begin{equation}
	x\frac{d^2F}{dx^2}+\frac{dF}{dx}-xF =0, \label{the_equation}
\end{equation}
where $x = \gamma r$. The solution of (\ref{the_equation}) converging at $x\rightarrow\infty$ is $F(r) = F_0 K_0(\gamma r)$, where $K_0$ is the Bessel-K function of the zero-th order. From the identity $dK_0(x)/dx =-K_1(x)$, where $K_1$ is the Bessel-K function of the first order, one finds a solution  for the electric potential and radial current density:
\begin{align}
	\phi(r,z) &= \phi_0 K_0(\gamma r)\cos{(\gamma z)} \label{phi_sol}\\
	j_r(r,z) &= \phi_0\gamma\sigma K_1(\gamma r)\cos{(\gamma z)} \label{jr_sol},
	\end{align}
where $\phi_0 =F_0G_0 $ is a constant. The resistance $R_c$ of the contact pad depicted in Fig.~\ref{fig1SM} is calculated easily for $\gamma t\ll1$. The total current flowing into the contact equals $I =\pi r_0t\cdot j_r(r=r_0)$, so that:
\begin{equation}
	R_c = \frac{1}{\pi\sigma t }\frac{K_0(\gamma r_0)}{\gamma r_0K_1(\gamma r_0)}, \label{contact_general}
\end{equation}
with $r_0=w/2$, where $w$ is the width of a wire to which the contact pad is attached. Using that $K_0(x)/xK_1(x)\approx \ln{(1.12/x)}$ in the limit of $x\rightarrow0$, one  can simplify the Eq.~(\ref{contact_general}) for $\gamma w\ll 1$: 

\begin{equation}
	R_c \approx \frac{\rho}{\pi t}\ln{\left(\frac{2.24}{\gamma w}\right)} \label{contact_approx}
\end{equation}

The Eq.~(\ref{contact_approx}) shows that in the case of poor interface ($\gamma w\ll1$) the contact resistance scales proportional to the material's resistivity $\rho$. The dependence of $\gamma$ on $\sigma$, see Eq.~(\ref{gamma}), is has a minor effect and enters via the log-term. Note that (\ref{contact_approx}) is very close to the qualitative estimate of (\ref{qualitative}) with $\lambda \approx \gamma^{-1}$.

\section{Solution: finite horn-shaped pad}

\begin{figure}[t]
	\includegraphics[width = 0.55\columnwidth]{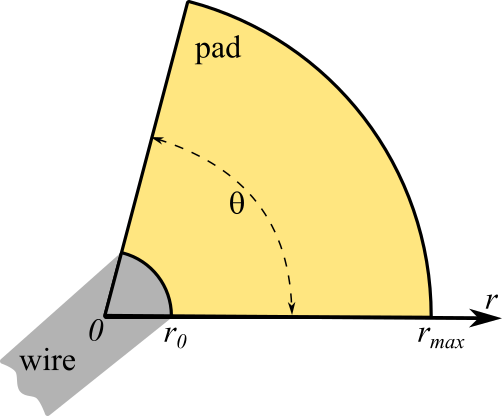}
	\caption{Sketch of a realistic contact pad, shaped as a horn with the inner radius of $r_0$ and the outer radius of $r_{\text{max}}$. The color notations are the same as in Fig.~\ref{fig1SM}.}
	\label{fig2SM}
\end{figure}

Consider a more realistic geometry of the finite size contact pad that has a shape of a horn. The angular width of the horn is $\theta$ and the inner and outer radii are, respectively, $r_0$ and $r_{\text{max}}$, see the Fig.~\ref{fig2SM}. In this case the solution of the Eq.~(\ref{the_equation}) is found from a different boundary condition, namely from $j_r(r=r_{\text{max}})=0$, i.e. zero in-plane current at the outer biundary of the lower layer. In this case $G(z)$ is the same and $F(r)$ becomes a superposition of the two Bessel functions. The expressions for the electric potential and current density are: 

\begin{align}
	\frac{\phi(r,z)}{\phi_0} &= \cos{(\gamma z)}\left[K_0(\gamma r)+\frac{K_1(\gamma r_{\text{max}})}{I_1(\gamma r_{\text{max}})}I_0(\gamma r)\right]\label{phi_sol_horn}\\
	\frac{j_r(r,z)}{\phi_0\gamma\sigma} &=  \cos{(\gamma z)}\left[K_1(\gamma r)-\frac{K_1(\gamma r_{\text{max}})}{I_1(\gamma r_{\text{max}})}I_1(\gamma r)\right] \label{jr_sol_horn},
	\end{align}
where $I_0$ and $I_1$ are the Bessel-I functions of the zero-th and first order. In the limit of $r_{\text{max}}\gg \gamma^{-1}$ the solution coincides with that for an infinite contact pad. The resistance of the contact pad of Fig.~\ref{fig2SM} is given by: 
\begin{equation}
	R_c = \frac{1}{\theta\sigma t x_0}\frac{K_0(x_0)I_1(x_{\text{max}})+I_0(x_0)K_1(x_{\text{max}})}{K_1(x_0)I_1(x_{\text{max}})-I_1(x_0)K_1(x_{\text{max}})}, \label{horn_sol}
\end{equation}
where $x_0=\gamma r_0$ and $x_{\text{max}}=\gamma r_{\text{max}}$ and $\theta$ in the denominator takes into account the angular width of the horn. For $x_0\ll x_{\text{max}}$, $K_1(x_0)I_1(x_{\text{max}})\gg I_1(x_0)K_1(x_{\text{max}})$ and the Eq.~(\ref{horn_sol}) simplifies:
\begin{equation}
	R_c = \frac{1}{\theta\sigma t x_0}\left[\frac{K_0(x_0)}{K_1(x_0)}+\frac{I_0(x_0)K_1(x_{\text{max}})}{K_1(x_0)I_1(x_{\text{max}})}\right]\label{horn_sol_simpl}
\end{equation}

For $x_{\text{max}}\gg1$ the contact pad size is much larger than $\gamma^{-1}$ and the second term in the brackets vanishes exponentially. In this case the result for $R_c$ reproduces (\ref{contact_general}) with $\theta$ instead of $\pi$ in the denominator, which is a result of reduced angular width. More interesting is the case the pad size smaller than the current transfer length, $x_0\ll x_{\text{max}}<1$. Using the asymptotic  approximations $I_0(x_0)\approx1$, $K_0(x_0)\approx\ln{(1.12/x_0)}$, $K_1(x_0)\approx 1/x_0$, $I_1(x_{\text{max}})\approx x_{\text{max}}/2$ and $K_1(x_{\text{max}})\approx 1/x_{\text{max}}$, we find:
\begin{equation}
	R_c = \frac{\rho}{\theta t}\ln{\left(\frac{1.12}{\gamma r_0}\right)}+\frac{1}{A_c\sigma_i},\, \text{with } A_c=\theta r_{\text{max}}^2/2\label{horn_final}
\end{equation}

Expression (\ref{horn_final}) shows that the resistance of a contact pad with $r_0\ll r_{\text{max}}<\gamma^{-1}$ has an extra contribution equal to the total interface resistance of the pad with the area of $A_c$. It is easy to see, that for $\sigma\rightarrow\infty$ ($\rho\rightarrow 0$) one has $\gamma\rightarrow0$ and (\ref{horn_final}) reduces to $R_c = (A_c\sigma_i)^{-1}$, that is the contact resistance between two ideal conductors is controlled solely by the interface resistance.

\section{Numerical modeling}

Finite interface resistance could explain substantial deviation in resistivity of two nanowire devices in Ref. [Liyang Chen et al., Science \textbf{382}, 907 (2023)]. The resistivity of the 0.66~$\mu$m long  wire is roughly an order of magnitude higher, as compared to the 28~$\mu$m long wire. Based on their response to the temperature and magnetic field similar to those of the parent film it is argued that this deviation is not a result of the contact resistance. The above calculations indicate that this argument has a caveat. 

As follows from (\ref{contact_approx}), the resistance of an infinite contact pad with a poor interface between the layers has nearly identical to the parent material temperature ($T$) response, since $R_c\propto\rho$ up to an unimportant log-factor. The same is expected to hold also for the magnetic field response. For a finite pad a constant resistance offset is present, according to (\ref{horn_final}). We have performed numerical calculations of realistic device layouts in order to understand possible effects of the interface conductivity and contact area on the temperature dependence of the resistivity in the smaller nanowire. 

\begin{figure}[t]
	\includegraphics[width = 0.79\columnwidth]{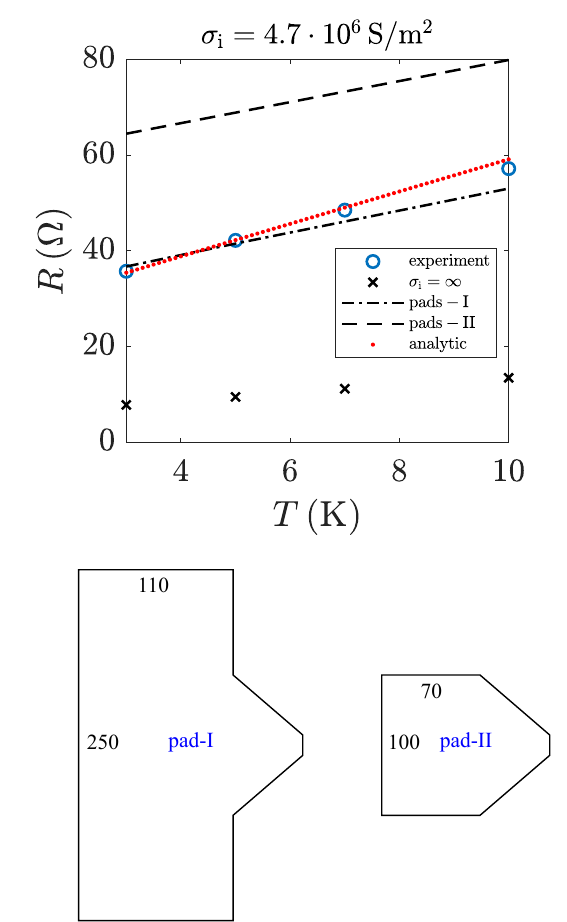}
	\caption{(upper panel): $T$-dependence of the resistance in a smaller wire. Circles mark the measured resistance and crosses correspond to $R_0=\rho L/(wt)$, expected from the resistivity measured in the 28~$\mu$m long wire under the assumption of ideal interface ($\sigma_i=\infty$). The theoretical fits take into account the contact resistance via $R = R_0+2R_c$. Red dotted line is the analytical result (\ref{horn_sol}) for $r_0=w/2$, $r_{\text{max}}=500\,\mu$m and $\theta=\pi/2$. Black lines are numerical simulations for the contact pads designs pads-I and pads-II. (lower  panel): sketches of the designs used in numerics.}
	\label{fig3SM}
\end{figure}

Fig.~\ref{fig3SM} (upper panel) shows the $T$-dependence of the resistance of the smaller wire with dimensions $L=0.66$~$\mu$m (length),  $w=0.24$~$\mu$m (width) and  $t=60$~nm (film thickness). Circles mark the measured resistance and crosses correspond to $R_0=\rho L/(wt)$, expected from the resistivity measured in the 28~$\mu$m long wire under the assumption of ideal interface ($\sigma_i=\infty$). This data is obtained by digitizing the published data. In order to capture the experimental behavior of the smaller wire we performed numerical simulations for two sizes of the contact pads, shaped as close as possible to the original design. The designs of a bigger (pads-I) and smaller (pads-II) contact pads are shown in the lower panel of Fig.~\ref{fig3SM}. Numerical results are shown in Fig.~\ref{fig3SM} by dashed-dotted and dashed black lines, respectively, both obtained assuming the interface conductance of  $\sigma_i=4.7\cdot10^6$\,S/m. The values of $\sigma_i,\rho$ and $t$ correspond to the current transfer length of $\gamma^{-1}\approx280\,\mu$m, which is comparable to the size of pads-I and considerably larger than the size of pads-II. The total device resistance $R=R_0+2R_c$ is much larger than $R_0$ (crosses) and exhibits sizable $T$-dependence, comparable to the experiment (circles). The resistance is larger for pads-II, as expected for their smaller area according to (\ref{horn_final}). Moreover, the experiment is consistent with the prediction of (\ref{horn_sol}) for $r_0=w/2$, $r_{\text{max}}=500\,\mu$m and $\theta=\pi/2$, see the red dotted line. Altogether, we find that poor interface quality can well be responsible for the unexpectedly high resistance of the smaller device. 

\section{Summary}

We observed that in the two-dimensional case the resistance of a contact pad caused by poor interface quality scales nearly proportional to the resistivity of a parent film. This could explain deviations in resistivity values between the larger and smaller devices in the experiments of Ref. [Liyang Chen et al., Science \textbf{382}, 907 (2023)]. The observation has important consequence for the interpretation of shot noise data in a smaller wire, since the effective length of the device may exceed the electron-phonon relaxation length. In this situation the observed reduction of the shot noise may not be surprising.
\end{document}